\documentclass[9pt,twocolumn,twoside]{opticajnl}
\journal{opticajournal} 

\setboolean{shortarticle}{true}

\usepackage{empheq}
\usepackage{textcomp}
\usepackage{bigints}
\usepackage{colortbl}
\usepackage{multirow}
\usepackage{array}
\usepackage{graphicx}
\usepackage{braket}
\usepackage{float}
\usepackage{bigints}
\usepackage{makecell}
\usepackage{graphicx}
\usepackage{dcolumn}
\usepackage{bm}
\usepackage{hyperref}
\usepackage[capitalise]{cleveref}
\usepackage{siunitx}
\usepackage{graphicx} 
\usepackage{booktabs}
\usepackage{upgreek}

\usepackage{pdfpages}

\usepackage[utf8]{inputenc}
\usepackage[T1]{fontenc}

\title{Mapping Nonlinear Mode Interactions in Coupled Kerr Resonators}
\author[1]{Luca O. Trinchão}
\author[1]{Luiz Peres}
\author[1]{Eduardo S. Gonçalves}
\author[1]{Miguel Nienstedt}
\author[2]{Laís Fujii dos Santos}
\author[1]{Paulo F. Jarschel}
\author[1]{Thiago P. M. Alegre}
\author[3]{Nathalia B. Tomazio}
\author[1,*]{Gustavo S. Wiederhecker}

\affil[1]{Gleb Wataghin Institute of Physics, University of Campinas, Campinas, SP, Brazil}
\affil[2]{School of Electrical Engineering and Computer Science, University of Ottawa, Ottawa, ON, Canada}
\affil[3]{Instituto de Física, Universidade de São Paulo, São Paulo, SP, Brazil}

\affil[*]{gsw@unicamp.br}

\begin{abstract}
We present a method for resolving spatial mode overlaps in coupled microresonators based on Kerr and thermal cross-phase modulation.
Through a pump-probe setup, we measure experimental overlap in a three-ring resonator with good agreement with analytical theory.
Our technique can be generalized for describing nonlinear interactions in more complex multi- and coupled-mode systems.

\end{abstract}

\setboolean{displaycopyright}{false} 

\begin{document}

\maketitle

\section{Introduction}

\begin{figure}[ht!]
\centering
\includegraphics[width=\linewidth]{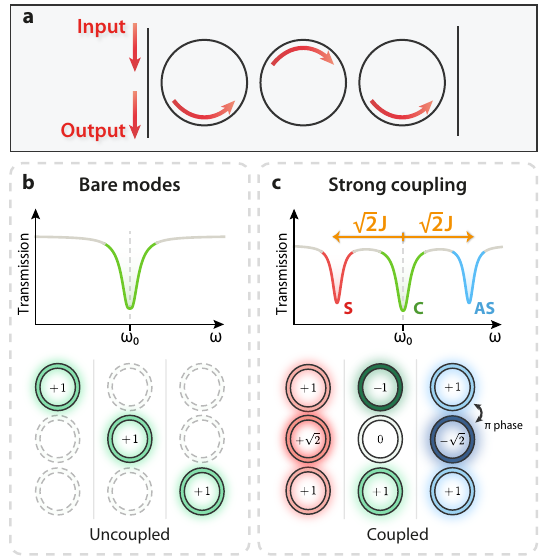}
\caption{\label{fig:1}
\justifying
\textbf{(a)} Chain of three coupled microresonators.
\textbf{(b)} 
Uncoupled (bare-ring) schematic representation.
Top: degenerate resonance frequencies of the uncoupled microrings. Bottom: uncoupled optical profiles showing the electric field amplitude in each resonator.
\textbf{(c)} Frequency-split supermodes under strong coupling: symmetric (S, red), central (C, green), and anti-symmetric (AS, blue).
Top: optical resonances corresponding to the hybridized supermodes. Bottom: supermode field profiles projected onto the bare-ring basis (unnormalized). “0” signifies non-resonant excitation, and negative amplitudes with dark shading indicate a relative $\pi$ phase difference.}
\end{figure}

Kerr microresonators are one of the fundamental blocks of nonlinear photonics, combining high intracavity power and small footprint to achieve strong nonlinear interactions at low pumping powers~\cite{del2007optical, herr2016dissipative, gaeta2019photonic, pal2025hybrid}. 
They span applications ranging from telecommunication~\cite{marin2017microresonator, kemal2016multi} and photonic computing~\cite{aghaee2025scaling, madsen2022quantum, okawachi2020demonstration, ghosh2024phase, wanjura2024fully, biasi2024exploring} to quantum information~\cite{larsen2025integrated, dutt2015chip} and spectroscopy~\cite{dutt2018chip, yu2017microresonator}.
Coupled resonator systems (\cref{fig:1}) expand this concept by enabling multimode excitation across adjacent resonators, leading to mode hybridization and allowing spectral engineering~\cite{haus2002coupled}.
Two- and three-ring architectures have been explored for high-purity quantum information processing~\cite{zhang2021squeezed, liao2020photonic, li2020photon}, high-efficiency wavelength conversion~\cite{tomazio2024tunable, gentry2014tunable}, and dispersion-engineered microcombs~\cite{xue2015normal, gao2024observation, ji2025multicolor, sanyal2025nonlinear}.

\begin{figure*}[ht!]
\centering
\includegraphics[width=.87\textwidth]{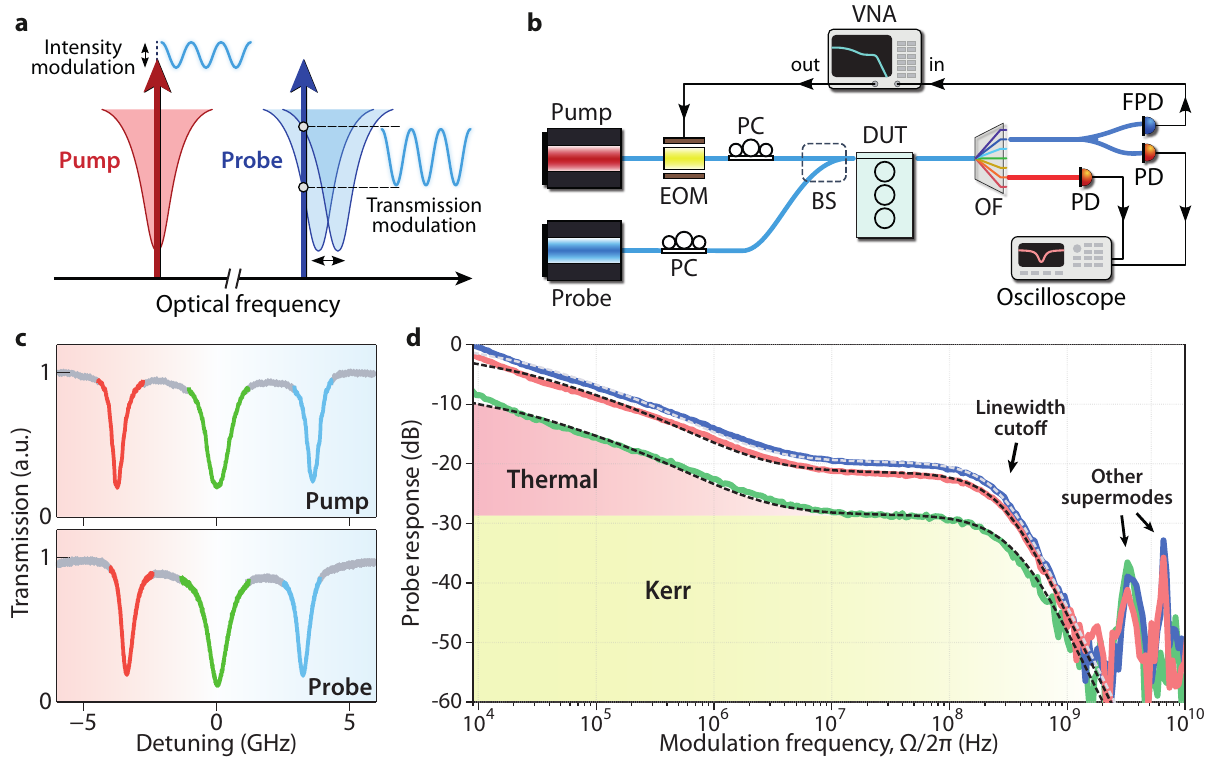}
\caption{\label{fig:2}
\justifying
\textbf{(a)} Schematic of the mode overlap experiment. 
An intensity-modulated pump ($\lambda_\mathrm{p} \approx 1546$~nm), resonant with a supermode, induces cross-phase modulation on a probe ($\lambda_\mathrm{b} \approx 1542$~nm) via Kerr and thermo-optic effects.
This leads to a modulation of the probe’s resonance frequency and, consequently, its transmission.
\textbf{(b)} Experimental setup for the overlap characterization. EOM electro-optic modulator, PC polarization controller, BS 50:50 beam splitter, DUT device under test, OF optical filter, PD photodetector, FPD fast photodetector, VNA vector network analyzer.
\textbf{(c)} Spectrum of the pumped and probed TE-polarized supermode triplets, centered at $\lambda_p$ and $\lambda_b$, respectively. S, C, and AS resonances are indicated in red, green, and blue. 
Measured loaded Q-factors are approximately $3.5 \times 10^5$ for the S and AS modes, and $2 \times 10^5$ for the C mode.
\textbf{(d)} Probe AC transmission versus modulation frequency, with the pump fixed at the S supermode and the probe set into S (red), C (green), and AS (blue). 
Shaded regions highlight dominant Kerr (yellow) and thermal (red) contributions. Dashed lines are fits from the theoretical model.
}
\end{figure*}

These systems exploit near-field evanescent coupling to produce hybridized, frequency-split coupled modes (supermodes) with tunable group velocity dispersion, unlocking customization of phase-matching in four-wave mixing (FWM) processes~\cite{tomazio2024tunable, gentry2014tunable, xue2015normal, miller2015tunable}.
While dispersion control has been mainly studied~\cite{tomazio2024tunable, gentry2014tunable, miller2015tunable, xue2015normal, souza2015spectral, mehrabad2025multi, flower2024observation}, the spatial structure of these modes, especially in systems with three or more resonators, is often overlooked. 
In dimer designs, coupling leads to two eigenstates -- the symmetric and anti-symmetric supermodes -- with similar field distribution profiles that differ only in phase~\cite{haus2002coupled, pal2024linear, hashemi2022uniform}.
On the other hand, higher-order systems support supermodes with varied spatial profiles, including bright and quasi-dark states~\cite{tomazio2024tunable,ghosh2024controlled, zeng2014design}.
These profiles strongly affect the strength of Kerr interactions through the spatial overlap of participating modes, modifying both self- and cross-phase modulation (SPM and XPM)~\cite{zeng2014design}, as well as the thermo-optic response, due to asymmetric heating across the resonators.
If left unaccounted, these effects can limit the efficiency and narrow the stable parameter window of nonlinear interactions such as microcomb formation, harmonic generation, and
squeezed-light sources~\cite{tomazio2024tunable, zeng2014design, sanyal2025nonlinear, zhang2019sub, soares2023third, vorobyev2025optimization, tatarinova2025optimization}.

In this work, we experimentally resolve supermode overlaps in a silicon nitride three-ring system (\cref{fig:1}(a)) using a pump-probe configuration.
By analyzing the Kerr and thermal cross-phase modulation response, we extract mode overlap values with excellent agreement with theoretical predictions.
Three resonators form the minimal set that supports distinct spatial mode profiles, though our analysis can be extended to larger arrays and more complex multimode architectures.

\section{Overlap analysis}

We investigate a photonic molecule composed of three identical evanescently coupled microring resonators (\cref{fig:1}(a)). Linear coupling gives rise to three hybridized supermodes, hereafter referred to as symmetric (S), central (C), and anti-symmetric (AS)~\cite{tomazio2024tunable}, with $\omega_\mathrm{S}<\omega_\mathrm{C}<\omega_\mathrm{AS}$ (\cref{fig:1}(c, top)). 
The frequency splitting between modes is given by $\sqrt{2}J$, where $J$ is the coupling rate, typically on the order of a few GHz.
In the spatial domain, S and AS are bright modes distributed across all three rings, whereas C is a quasi-dark mode with a non-resonant field in ring 2~\cite{souza2016modeling}, acting only as a waveguide between rings 1 and 3 (\cref{fig:1}(c, bottom)).

Due to their distinct spatial profiles, the strength of Kerr XPM between supermodes is proportional to their overlap factor~\cite{chembo2010modal}:
\begin{equation}
    \Gamma_{p,b} = \frac{1}{4}\, \bigintssss_V \epsilon^2 \left| \vec{\Upsilon}_b^* \cdot \vec{\Upsilon}_p \right|^2\, dV.\label{eq:overlap}
\end{equation}
over the three-ring volumes ($V$). Here, $\vec{\Upsilon}_{p(b)}$ denotes the complex electric-field spatial profile of the pump supermode (probe), normalized such that \cref{eq:overlap} has units of $[1/\mathrm{m}^3]$ (see Supplementary Material, S3). In \cref{fig:1}(c, bottom), $\vec{\Upsilon}_\mathrm{S}$, $\vec{\Upsilon}_\mathrm{C}$, and $\vec{\Upsilon}_\mathrm{AS}$ are schematically represented.


Inspired by the approach of Gao \textit{et al.} for probing absorption and optical nonlinearities in integrated microresonators~\cite{gao2022probing}, we exploit XPM induced by Kerr and thermal effects to experimentally access the spatial overlap between supermodes in a pump-probe experiment, described below.
A pump laser ($\lambda_\mathrm{p} = 1546$~nm), intensity-modulated at frequency $\Omega$, is injected into a supermode resonance. 
A co-propagating probe ($\lambda_\mathrm{b} = 1542$~nm) is placed one free spectral range (FSR) away from the pump, blue-detuned from a supermode, allowing for high-excitation filtering of pump and probe channels.
In this setup, the probe undergoes changes in amplitude and phase, as the ring cavity experiences effective refractive index modulation as a result of the joint effects of instantaneous Kerr and slower thermal XPMs. 
This modulates the probe's resonance frequency and thus its transmission (\cref{fig:2}(a)).

We implement the setup shown in \cref{fig:2}(b), scanning the modulation frequency from \qty{9}{\kilo\hertz} to \qty{20}{\giga\hertz} using a vector network analyzer (VNA), which simultaneously records the probe’s AC response. 
Optical powers are kept low to avoid optical bistability.
Maximum transduction is observed for probe detunings in the range of 0.2-0.3 linewidths, consistent with the detuning that maximizes the transmission slope 
$\bigl| dT_b/d\Delta \bigr|_{\mathrm{max}}$ at $\Delta = \kappa/(2\sqrt{3})$.
The supermode triplets studied are shown in \cref{fig:2}(c).
As the Kerr effect is virtually instantaneous, while the thermal response occurs on a slower ($\approx \unit{\micro\second}$) timescale, their contributions can be spectrally separated~\cite{gao2022probing, rokhsari2005observation}.

We characterize the mode overlap by tuning the pump into a supermode in one triplet and probing the three supermodes in another. \cref{fig:2}(d) shows the probe response when pumping the S mode (responses for pumping other supermodes are provided in the Supplementary Material, Fig.~S2). 
At low modulation frequencies, both Kerr and thermal effects contribute. 
As the frequency increases, the slower thermal response is progressively suppressed, leaving the Kerr effect as the dominant mechanism.
At higher frequencies, the modulation sidebands fall outside the optical linewidth of the cavity, and the response vanishes.
However, when the modulation frequency matches the optical frequency splitting of adjacent supermodes ($\sim$\qty{3.5} or \qty{7}{\giga\hertz}), a modulation sideband falls back into resonance again, resulting in the peaks seen in the experimental traces above \qty{2}{\giga\hertz}.

\begin{figure}
    \centering
    \includegraphics[width=\linewidth]{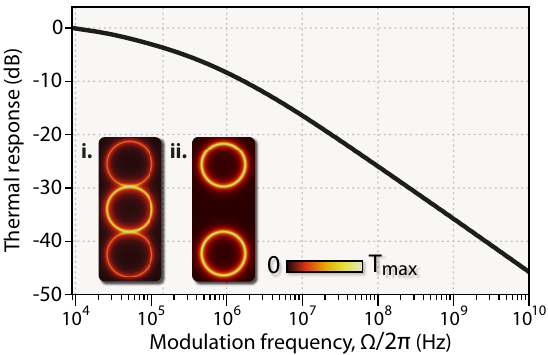}
    \caption{
    \justifying
    Thermal response of the S, C, and AS supermodes as a function of the modulation frequency. Inset: Temperature profile of the modulated absorbed power for the S/AS (i.) and C (ii.) supermodes, simulated in COMSOL Multiphysics with the \textit{Heat Transfer in Solids} module.}
    \label{fig:3}
\end{figure}

At low modulation frequencies, the probe frequency shift is dominated by thermal XPM, which can be expressed as~\cite{gao2022probing}:
\begin{equation}
\widetilde{\delta_{th}}(\Omega) = G_{p,b}^{\theta} \, \tilde{r}(\Omega) \, \widetilde{I_p}(\Omega).\label{eq:response}
\end{equation}
Here, $\widetilde{I_p}(\Omega)$ [\si{\joule}] is the modulated intracavity pump energy, $\tilde{r}(\Omega)$ is the dimensionless thermal relaxation response, and $G_{p,b}^{\theta}$ [\si{\hertz/\joule}] is the thermal frequency pulling, describing the probe frequency shift per unit of absorbed pump energy~\cite{primo2021accurate}.
As the pump-induced temperature distribution and probe optical profile do not coincide perfectly, $G_{p,b}^{\theta}$ is obtained by integrating the thermo-optic effect across the volume of the three rings (see Supplementary Material, Eq. S35).

We study the thermal dynamics by simulating the response of the device with a 3D finite element model (\cref{fig:3})~\cite{gao2022probing}.
In the simulation, the heat source follows the spatial profile of the pump's intracavity energy and oscillates in time to produce the thermal response shown in \cref{fig:3}.
For each probed supermode, the harmonic temperature component is integrated over the SiN ring domains, where the nonlinearity is strongest and the optical mode is primarily confined.
\cref{fig:3} shows the normalized thermal response $\tilde{r}(\Omega)$ and the temperature profiles for the S/AS and C supermodes under absorption (inset). 
The thermal response was found to be identical for all supermode excitations, indicating that each individual ring experiences a similar thermal relaxation process.
Moreover, since S and AS differ only in phase, their thermal profiles are identical, as seen in \cref{fig:3}(i.).


As shown in \cref{fig:2}(d), the probe response is strongest when the spatial overlap with the pump mode is high.
Because the S and AS supermodes share similar mode profiles (see \cref{fig:1}(c)), they exhibit efficient cross-phase modulation.
In contrast, the C supermode (green) has reduced field overlap with the others, resulting in a weaker response. 
The dashed lines in \cref{fig:2}(d) are fits to a theoretical model that incorporates the low-frequency thermal response together with the pump and probe linewidth cutoffs (details in the Supplementary Material, Eq.~S39).
From these fits, we extract relative mode overlaps between pumped and probed supermodes. 

For ease of comparison between pump choices, we report a dimensionless proxy, $\eta_{p,b}\equiv \Gamma_{p,b}/{\Gamma_{p,p}}$,
which rescales the unnormalized overlaps while preserving their relative order.
These values are summarized in \cref{tab:results}, along with theoretical predictions based on the analytical eigenvectors of the coupled modes (\cref{fig:1}(c, bottom)). 
While \cref{fig:2}(d) illustrates the probe response for a single pumped supermode, evaluating all overlaps listed in \cref{tab:results} requires probing all nine combinations of pumped and probed supermodes, summarized in the Supplementary Material, Fig.~S2 to prevent redundancy in the main text.
The experimental and theoretical values are in good agreement, with minor deviations for the AS mode.
We attribute these to residual cavity frequency mismatches that slightly distort the supermode profiles, despite our best efforts to compensate for these shifts through the integrated microheaters that control the temperature of each individual ring.

\begin{table}[t!]
\centering
\setlength{\tabcolsep}{5pt}
\renewcommand{\arraystretch}{1.1}
\begin{tabular}{lccc}
\hline
& \textbf{S} & \textbf{C} & \textbf{AS} \\
\hline
\multicolumn{4}{l}{\textbf{Theoretical}} \\
Overlap with S   & 1.00 & 0.67 & 1.00 \\
Overlap with C   & 0.50 & 1.00 & 0.50 \\
Overlap with AS  & 1.00 & 0.67 & 1.00 \\
\hline
\multicolumn{4}{l}{\textbf{Experimental}} \\
Overlap with S   & $1.00$ & $0.69 \pm 0.03$ & $1.00 \pm 0.02$ \\
Overlap with C   & $0.49 \pm 0.02$ & $1.00$ & $0.52 \pm 0.01$ \\
Overlap with AS  & $1.10 \pm 0.03$ & $0.81 \pm 0.03$ & $1.00$ \\
\hline
\end{tabular}
\caption{Relative mode overlap ($\Gamma_{p,b}/\Gamma_{p,p}$) between pumped and probed supermodes. Diagonal entries correspond to the self-overlap of each supermode, normalized to unity by definition. 
Values exceeding one indicate XPM between different supermodes is stronger than corresponding self-interactions (across the same supermode), reflecting deviations of the supermode structures.
}
\label{tab:results}
\end{table}

\section{Conclusions}
In summary, we have investigated the role of nonuniform XPM in a three-microring system.
We introduce a pump-probe method that exploits the different timescales of Kerr and thermal XPM to extract mode overlaps consistent with theoretical predictions.
Our framework generalizes naturally to larger coupled-resonator networks and to a wide range of quadratic and cubic nonlinear effects.
These results advance the understanding and modeling of nonlinear processes in coupled microresonators and their applications in coherent computing and quantum photonics.

\begin{backmatter}
\bmsection{Funding}
This work was supported by São Paulo Research Foundation (FAPESP) through grants 
18/15577-5, 
18/15580-6, 
18/25339-4, 
21/10334-0, 
23/09412-1, 
24/15935-0, 
25/04049-1, 
20/04686-8, 
22/06267-8, 
24/02289-2, 
24/14425-8, 
18/21311-8, 
24/04845-0  
23/12741-7 
and Coordenação de Aperfeiçoamento de Pessoal de Nível Superior - Brasil (CAPES) (Finance Code 001)

\bmsection{Acknowledgment} 
The authors thank André Primo for insightful discussions on the thermal dynamics of microresonators

\bmsection{Disclosures} The authors declare no conflicts of interest.

\bmsection{Data availability} Data underlying the results of this paper will be made available in Zenodo upon publication (DOI to be provided).

\bmsection{Supplemental document}
The Supplementary Material provides additional details on the three-ring resonator model, device specifications, and the probe response model used to extract the mode overlaps.

\end{backmatter}

\bibliography{bib}



\end{document}


\maketitle

\section{Three-ring resonator model}

For a system of three identical evanescently coupled microring resonators, the mode hybridization is governed by the coupling matrix~\cite{tomazio2024tunable}:
\begin{equation}
    \mathbb{H} =
    \begin{pmatrix}
        \omega_0 & -J & 0 \\
        -J & \omega_0 & -J \\
        0 & -J& \omega_0
    \end{pmatrix}
    ,
    \label{eq:Hmatrix}
\end{equation}
where $\omega_0$ is the resonance frequency of the uncoupled rings and $J$ is the linear coupling rate between adjacent resonators. 

Diagonalization of $\mathbb{H}$ yields the eigenfrequencies and eigenvectors that describe the hybridized supermodes:
\begin{subequations}
	\begin{align}
	&\vec{b}_\mathrm{S} = \frac{1}{2}(1~\sqrt{2}~1)^T, & &\omega_\mathrm{S} = \omega_0 - \sqrt{2}J. & \\
	&\vec{b}_\mathrm{C} = \frac{1}{\sqrt{2}}(-1~0~1)^T, & &\omega_\mathrm{C} = \omega_0, &  \\
	&\vec{b}_\mathrm{AS} = \frac{1}{2}(1~-\sqrt{2}~1)^T, & &\omega_{\mathrm{AS}} = \omega_0 + \sqrt{2}J, &  
	\end{align}
\end{subequations}
corresponding to the symmetric (S), central (C), and anti-symmetric (AS) supermodes, respectively, with $\omega_\mathrm{S} < \omega_\mathrm{C} < \omega_\mathrm{AS}$.

The eigenvectors $\{\vec{b}_\mathrm{S}, \vec{b}_\mathrm{C}, \vec{b}_\mathrm{AS}\}$ describe the supermode field distributions as projected onto the bare-ring basis, where the $j$-th component indicates the relative field amplitude in ring $j$. As an example, the optical mode profile of the symmetric supermode can be expressed as:
\begin{equation}
    \vec{\Upsilon}_{\mathrm{S}} = \vec{b}_{\mathrm{S}} \cdot
    \begin{pmatrix}
        \vec{\Upsilon}_1 \\
        \vec{\Upsilon}_2 \\
        \vec{\Upsilon}_3
    \end{pmatrix}
    = \tfrac{1}{2} \vec{\Upsilon}_1 + \tfrac{\sqrt{2}}{2} \vec{\Upsilon}_2 + \tfrac{1}{2}\vec{\Upsilon}_3,
\end{equation}
where $\vec{\Upsilon}_j$ denotes the optical field profile of the $j$-th ring.

The spatial distributions of all three supermodes are illustrated in Fig.~1(c) of the main text.

\section{Device specifications}

The device consists of three \qty{50}{\micro\meter}-radius silicon nitride microring resonators (\qty{1}{\micro\meter} width $\times$ \qty{0.8}{\micro\meter} height) cladded in silica. Inter-ring gaps of \qty{800}{\nano\meter} yields frequency splitting of approximetely \qty{3.5}{\giga\hertz} at \qty{1550}{\nano\meter}, while a bus-waveguide gap of \qty{600}{\nano\meter} results in undercoupling. Integrated microheaters on top of each ring are incorporated for frequency tunability via thermo-optic control. Representative optical transmissions are shown in Fig.~2(c) of the main text.

\section{Model for mode overlap characterization}\label{sec:SM_charac}
Here, we describe the model used for the pump-probe experiment to characterize the relative mode overlap of hybridized supermodes in the three-microring coupled system.
This approach builds upon the framework introduced by Gao \textit{et al.}~\cite{gao2022probing} for probing absorption and nonlinear responses in integrated microresonators. 
We extend this technique by comparing the distinct responses obtained when pumping and probing different supermodes of the same device. Because these modes exhibit distinct hybridized profiles, this allows for a direct measurement of their relative spatial mode overlaps.

We begin by writing the electric field of a given resonant mode $\eta$ as:
\begin{equation}
    \vec{E}_\eta(\vec{r},t) = a_\eta(t)\vec{\Upsilon}_\eta(\vec{r})e^{i\omega_\eta t} + c.c.,
\end{equation}
where $a_\eta(t)$ is the slowly varying complex amplitude of the $\eta$ optical mode, normalized such that $|a_\eta|^2$ is the circulating energy stored in the optical mode, $\omega_\eta$ is the optical resonance frequency, and $\vec{\Upsilon}_\eta(\vec{r})$ is the spatial mode profile, normalized as
\begin{equation}
    \frac{1}{2} \int_{\infty} \epsilon(\vec{r}) \left|\vec{\Upsilon}_\eta(\vec{r})\right|^2 dV = 1.
\end{equation}

\subsection{Pump response}

We consider a pump mode $a_p(t)$ driven by an input field $s_p(t)$ that is composed of a steady-state component and a time-dependent modulation:
\begin{equation}
    s_p(t) = s_p^{(0)} + \varepsilon \,\delta s_p(t),
\end{equation}
where $s_p^{(0)}$ is the DC the input field, and $\delta s_p(t)$ is an AC perturbation.

For a pump laser is positioned near resonance ($\Delta_p \approx 0$, small detuning changes can be neglected to first order), the time-domain equation for $a_p(t)$ reads:
\begin{equation}
    \frac{da_p(t)}{dt} = -\frac{\kappa_p}{2}a_p(t) + \sqrt{\kappa_{p,ex}}s_p(t).,
    \label{eq:pumpeq}
\end{equation}
where $\kappa_p = \kappa_{p,in} + \kappa_{p,ex}$ is the pump total losses, with intrinsic loss $\kappa_{p,in}$ and coupling loss $\kappa_{p,ex}$ .

Given that $s_p(t)$ is modulated, $a_p(t)$ can also be expressed in terms of a DC and AC component:
\begin{equation}
    a_p(t) = a_p^{(0)} + \varepsilon\,\delta a_p(t).
    \label{eq:pumpexp}
\end{equation}

Inputting Eq.\ref{eq:pumpexp} into Eq.\ref{eq:pumpeq} and collecting the terms of order $\varepsilon^0$, we obtain the steady-state solution:
\begin{equation}
    a_p^{(0)} = \frac{\sqrt{\kappa_{p,ex}}}{\kappa_p/2}s_p^{(0)}.
\end{equation}

The AC equation for $\delta a_p(t)$ is found by collecting the terms of order $\varepsilon^{1}$:
\begin{equation}
    \delta a_p'(t) = -\frac{\kappa_p}{2} \delta a_p(t) + \sqrt{\kappa_{p,\mathrm{ex}}}\, \delta s_p(t).
\end{equation}
where the LHS of the equation indicates the time derivative.

Taking the Fourier transform (\( \tilde{z}(\Omega) = \mathcal{F}\{z(t)\} \)), we find:
\begin{equation}
    \widetilde{\delta a_p}(\Omega) = \frac{\sqrt{\kappa_{p,\mathrm{ex}}}}{\kappa_p/2 + i\Omega} \, \widetilde{\delta s_p}(\Omega),
\end{equation}
which describes the frequency-domain response of the intracavity field to the modulation frequency.

The oscillating intracavity energy $I_p(t) = |a_p(t)|^2$ has a Fourier component given by:
\begin{equation}
    \widetilde{I_p}(\Omega) \approx \left(a_p^{(0)}\right)^* \widetilde{\delta a_p} + a_p^{(0)} \widetilde{\delta a_p^*},
\end{equation}
and similarly, the modulated input power is:
\begin{equation}
    \widetilde{P_{\mathrm{in},p}}(\Omega) \approx \left(s_p^{(0)}\right)^* \widetilde{\delta s_p} + s_p^{(0)} \widetilde{\delta s_p^*}.
\end{equation}

So the normalized intracavity energy can be written as:
\begin{equation}
    \frac{\widetilde{I_p}}{\widetilde{P_{in,p}}} = \frac{2\eta_p}{\kappa_p/2+i\Omega},
    \label{eq:IpPin}
\end{equation}
where $\eta_p = \kappa_{p,ex}/\kappa_p$ is the pump mode coupling efficiency. 
\cref{eq:IpPin} describes how the $\widetilde{I_p}$ exhibits a low-pass filter response.
For modulation frequencies $\Omega>\kappa_p$, the modulation sidebands are filtered out by the cavity linewidth, and the intracavity field experiences only the DC component.

\subsection{Probe response}
We now consider a probe mode $a_b(t)$, which experiences modulation through a perturbation in its detuning $\Delta_b(t)$ due to the presence of a modulated pump:
\begin{equation}
    a_b(t) = a_b^{(0)} + \varepsilon\,\delta a_b(t), \quad \quad \Delta_b(t) = \Delta_b^{(0)} + \varepsilon\,\delta_b(t).
\end{equation}
$\delta_b(t)$ is a time-varying phase-shift due to thermal-optic and Kerr XPM from the pump. These mechanisms are detailed in a later section. The time evolution of the probe field is written as:
\begin{equation}
    \frac{da_b(t)}{dt} = -\left( \frac{\kappa_b}{2}+i\Delta_b(t)\right)a_b(t) + \sqrt{\kappa_{b,ex}}s_b.
\end{equation}

Collecting terms of order $\varepsilon^0$, we find:
\begin{equation}
    0 = -\left(\frac{\kappa_b}{2} + i\Delta_b^{(0)}\right)a_b^{(0)} + \sqrt{\kappa_{b,ex}}s_b
\end{equation}
which yields the DC solution:
\begin{equation}
    a_b^{(0)} = \frac{\sqrt{\kappa_{b,ex}}}{\kappa_b/2+i\Delta_b^{(0)}}s_b.
\end{equation}
Additionally, the AC equation reads:
\begin{equation}
    \delta a_b'(t) = -\left(\frac{\kappa_b}{2} + i\Delta_b^{(0)}\right) \delta a_b(t) - i \delta_b(t)a_b^{(0)}. 
\end{equation}
Taking the Fourier transform of this expression yields:
\begin{equation}
    \widetilde{\delta a_b}(\Omega) = \frac{-a_b^{(0)}}{\kappa_b/2 + i(\Delta_b^{(0)} + \Omega)}~\widetilde{\delta_b}(\Omega).
\end{equation}

The probe transmission is given by:
\begin{equation}
    T_b(t) = \left| s_b - \sqrt{\kappa_{b,ex}}\,a_b(t) \right|^2.
\end{equation}
Up to first order in $\varepsilon$, the AC component of the transmission is:
\begin{equation}
    \widetilde{T_b}(\Omega) = \left( s_b - \sqrt{\kappa_{b,ex}}\,a_b^{(0)} \right)\sqrt{\kappa_{b,ex}}\,\widetilde{\delta a_b^*}(\Omega)
    + \left( s_b^* - \sqrt{\kappa_{b,ex}}\,\left(a_b^{(0)}\right)^* \right)\sqrt{\kappa_{b,ex}}\,\widetilde{\delta a_b}(\Omega),
\end{equation}
where we discard DC ($\varepsilon^0$) and higher-order ($\varepsilon^n$, $n \geq 2$) contributions.

Substituting $a_b^{(0)}$ and $a_b(t)$, we arrive at:
\begin{equation}
    \frac{\widetilde{T_b}(\Omega)}{\widetilde{\delta_b}(\Omega)} = 
    - \frac{2\kappa_{b,ex}\,\Delta_b^{(0)}}{\left(\kappa_b/2\right)^2 + \left( \Delta_b^{(0)} \right)^2}
    \frac{\kappa_b - \kappa_{b,ex} + i \Omega}{\left(\kappa_b/2 + i\Omega\right)^2 + \left(\Delta_b^{(0)}\right)^2}
    P_{in,b}.
\end{equation}
This result describes the modulated response of the probe field driven by the pump-induced modulation in detuning $\widetilde{\delta_b}$.

\subsection{Thermal and Kerr XPMs}

We now focus on the probe’s modulated detuning term $\widetilde{\delta_b}(\Omega)$, which describes the AC phase-shift induced by Kerr and thermal XPM effects from the pump. It can be written as the sum of the two contributions:
\begin{equation}
    \widetilde{\delta_b}(\Omega) = \widetilde{\delta_{th}}(\Omega) + \widetilde{\delta_{\mathrm{NL}}}(\Omega),
\end{equation}
where $\widetilde{\delta_{th}}(\Omega)$ accounts for the thermal-optic phase shift, and $\widetilde{\delta_{\mathrm{NL}}}(\Omega)$ corresponds to the instantaneous Kerr-induced phase shift. We discuss each of these terms individually in the subsections below.

\subsubsection{Kerr XPM}
The Kerr XPM is given by:
\begin{equation}
    \widetilde{\delta_{\mathrm{NL}}}(\Omega) = -\gamma_{p,b}\, g_{0,b}\, \widetilde{I_p}(\Omega),
\end{equation}
where $g_{0,b}$ is the Kerr coefficient of the probe mode and $\gamma_{p,b}$ is the XPM factor that accounts for the spatial mode overlap between pump and probe. Because the Kerr response is virtually instantaneous, $\widetilde{\delta_{\mathrm{NL}}}$ dependency on the modulation frequency $\Omega$ arises only from $\widetilde{I_p}(\Omega)$.

The Kerr coefficient is defined as:
\begin{equation}
    g_{0,b} = \frac{\omega_b c n_2}{n_0 n_g V_{\mathrm{eff},b}},
\end{equation}
where $c$ is the vacuum speed of light, $n_0$ and $n_g$ are the material and group indexes, and $V_{\mathrm{eff},b}$ is the effective mode volume of the probe field across the three rings:
\begin{equation}
    V_{\mathrm{eff},b} = \frac{4}{\bigintssss_{V} \epsilon^2||\vec{\Upsilon}_b ||^4dV}.
\end{equation}

The XPM factor $\gamma_{p,b}$ is written as:
\begin{equation}
    \gamma_{p,b} = 2 \Gamma_{p,b} V_{\mathrm{eff},b},
\end{equation}
with $\Gamma_{p,b}$ being the nonlinear overlap integral between pump and probe modes~\cite{chembo2010modal}:
\begin{equation}
    \Gamma_{p,b} = \frac{1}{4} \int_V \epsilon^2 \left| \vec{\Upsilon}_b^* \cdot \vec{\Upsilon}_p \right|^2 dV.
\end{equation}

To evaluate $\Gamma_{p,b}$ across different supermodes, we express each supermode profile as projected onto bare-ring (uncoupled) microring using the coupled basis $\{\vec{b}_\mathrm{S}, \vec{b}_\mathrm{C}, \vec{b}_\mathrm{AS}\}$. For all combinations of pumped and probed supermodes,  the spatial overlap is calculated as described by ref. \cite{tomazio2024tunable}. Normalizing by the self-overlap of the pumped mode, we obtain:
\begin{equation}
    \begin{pmatrix}
        1/\Gamma_{\mathrm{S},\mathrm{S}} & 0 & 0\\
        0 & 1/\Gamma_{\mathrm{C},\mathrm{C}} & 0\\
        0 & 0 & 1/\Gamma_{\mathrm{AS},\mathrm{AS}}
    \end{pmatrix}
    \cdot
    \begin{pmatrix}
        \Gamma_{\mathrm{S},\mathrm{S}} & \Gamma_{\mathrm{S},\mathrm{C}} & \Gamma_{\mathrm{S},\mathrm{AS}}\\
        \Gamma_{\mathrm{C},\mathrm{S}} & \Gamma_{\mathrm{C},\mathrm{C}} & \Gamma_{\mathrm{C},\mathrm{AS}}\\
        \Gamma_{\mathrm{AS},\mathrm{S}} & \Gamma_{\mathrm{AS},\mathrm{C}} & \Gamma_{\mathrm{AS},\mathrm{AS}}
    \end{pmatrix}
    =
    \begin{pmatrix}
        1 & 2/3 & 1\\
        1/2 & 1 & 1/2\\
        1 & 2/3 & 1
    \end{pmatrix}.
    \label{eq:overlaps}
\end{equation}
These correspond to the theoretical values presented in Table 1 of the main text.

\subsubsection{Thermal XPM}

The thermo-optic effect induces a refractive index perturbation given by $\Delta n_{th}(\vec{r},t) = \alpha_n \Delta T(\vec{r},t)$, where $\alpha_n = \partial n/\partial T$ is the thermo-optic coefficient and $\Delta T(\vec{r},t)$ is the time-dependent temperature field. Using first-order perturbation theory, the corresponding phase-shift on the probe mode is~\cite{joannopoulos2008molding, primo2021accurate}:

\begin{equation}
\begin{split}
    \delta_{th}(t) &= -\frac{\omega_b}{2} \frac{\int_V \! \Delta \epsilon(\vec{r},t) |\Upsilon_b(\vec{r})|^2 dV}{\int_V \! \epsilon(\vec{r}) |\Upsilon_b(\vec{r})|^2 dV} \\
    &= -\omega_b \frac{\int_V \! n_0 \alpha_n \Delta T(\vec{r},t) |\Upsilon_b(\vec{r})|^2 dV}{\int_V \! \epsilon(\vec{r}) |\Upsilon_b(\vec{r})|^2 dV}.
\end{split}
\end{equation}

We decompose the temperature field into a normalized spatial profile and a time-dependent amplitude:
\begin{equation}
    \Delta T(\vec{r},t) = \delta T(\vec{r}) \, \theta(t).
\end{equation}
Here, $\delta T(\vec{r})$ is the dimensionless temperature mode profile (normalized such that $max\{\delta T(\vec{r})\} = 1$), determined by the spatial distribution of absorbed power from the optical mode. The temporal amplitude $\theta(t)$, with units of K, governs the AC thermal dynamics, and it is defined as:
\begin{equation}
    \theta (t) = R_{th} \kappa_{abs}  |a(t)|^2,
    \label{eq:thermaltime}
\end{equation}
where $R_{th}$ is the material thermal resistance (K/W) and $\kappa_{abs}$ is the linear absorption rate (Hz).
Although \cref{eq:thermaltime} does not account for the thermal relaxation time and acts as if the temperature diffusion occurs instantly, we address this effect below. 

Transforming to the frequency domain, the AC component up to the first order becomes:
\begin{equation}
    \widetilde{\delta_{th}}(\Omega) = -\omega_b \frac{\int_V \! n_0 \alpha_n R_{th} \kappa_{abs} \delta T_p(\vec{r}) |\Upsilon_b(\vec{r})|^2 dV}{\int_V \! \epsilon(\vec{r}) |\Upsilon_b(\vec{r})|^2 dV} \cdot \widetilde{I_p}(\Omega),
\end{equation}
where $\delta T_p(\vec{r})$ is the temperature profile induced by absorption from the pump mode. We simulate these distributions using COMSOL Multiphysics (see Fig. 3 of the main text), modeling each ring as a uniform power source proportional to the circulating optical power of the given supermode. This approach neglects the transverse optical field profile.

We define the thermal frequency pulling~\cite{primo2021accurate}:
\begin{equation}
    G_{p,b}^{\theta} = -\omega_b \cdot  \frac{\int_V \! n_0 \alpha_n R_{th} \kappa_{abs} \delta T_p(\vec{r}) |\Upsilon_b(\vec{r})|^2 dV}{\int_V \! \epsilon(\vec{r}) |\Upsilon_b(\vec{r})|^2 dV},
\end{equation}
with units of Hz/J, which describes the probe frequency shift per unit of absorbed pump energy.

Additionally, the thermal relaxation dynamics are captured by the normalized thermal response function:
\begin{equation}
\tilde{r}(\Omega) = \frac{\widetilde{\theta}(\Omega)}{\widetilde{\theta}(0)},
\end{equation}

This function is obtained by introducing an AC-modulated heat source in COMSOL with the same spatial distribution as $\delta T_p(\vec{r})$, and computing the resulting AC temperature response as a function of the modulation frequency $\Omega$. The resulting curve is shown in Fig. 3 of the main text. We find that the $\tilde{r}(\Omega)$ response is approximately the same for all pumped supermodes.

Combining these definitions, the thermal phase-shift can be written as:
\begin{equation}
    \widetilde{\delta_{th}}(\Omega) = G_{p,b}^{\theta} \tilde{r}(\Omega) \widetilde{I_p}(\Omega).
\end{equation}

\subsection{Probe tranismission}
\begin{figure}[h]
    \centering
    \includegraphics[width=0.5\linewidth]{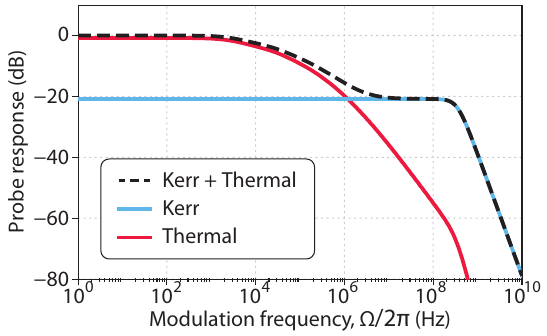}
    \caption{Probe response $\widetilde{T_b}$ as a function of the modulation frequency, highlighting the individual contributions of Kerr and thermal XPM. Black: both Kerr and thermal contributions included; blue: only Kerr; red: only thermal.}    
    \label{fig:Tb}
\end{figure}

The probe transmission can described by the combinations three effects: \textit{(i)} the build up of intracavity energy of the pump field, \textit{(ii)} the induced probe frequency shift due to Kerr and thermal XPM from the pump, and \textit{(iii)} the resulting modulation in the probe transmission~\cite{gao2022probing}:
\begin{equation}
    \frac{\widetilde{T_b}}{\widetilde{P_{in,p}}} = \frac{\widetilde{T_b}}{\widetilde{\delta_b}} \frac{\widetilde{\delta_b}}{\widetilde{I_p}} \frac{\widetilde{I_p}}{\widetilde{P_{in,p}}}.
    \label{eq:Tb}
\end{equation}
\Cref{fig:Tb} shows a typical transmission spectrum of $\widetilde{T_b}$ as a function of modulation frequency $\Omega$, highlighting the different Kerr and thermal contributions.

\subsection{Resolving the relative mode-overlap}
We perform the pump-probe experiment by alternating the pump and probe supermodes for all combinations. For each case, we fit the measured AC probe transmission $\widetilde{T_b}$ using the expression derived from \cref{eq:Tb}:
\begin{equation}
    \frac{\widetilde{T_b}}{\widetilde{P_{in,p}}} =
    N_{p,b} \cdot  \frac{\left(X_{p,b} + \tilde{r}(\Omega) \right)\Delta_b^{(0)}\kappa_{p,ex}\kappa_{b,ex}(\kappa_b-\kappa_{b,ex}+i\Omega)}{\kappa_p\left(4{\Delta_b^{(0)}}^2+\kappa_b^2\right)\left(2\Delta_b^{(0)}+i\kappa_b-2\Omega\right)\left(\kappa_p + 2i\Omega\right)\left(2\Delta_b^{(0)}-i\kappa_b+2\Omega\right)},
    \label{eq:Tb_redefined}
\end{equation}
where we defined a normalization constant:
\begin{equation}
    N_{p,b} = 128 P_{in,b}G_{p,b}^{\theta}
\end{equation}
and the Kerr-to-thermal XPM ratio:
\begin{equation}    
    X_{p,b} = \frac{g_{0,b} \gamma_{p,b}}{G_{p,b}^{\theta}}.
\end{equation}
In practice, $N_{p,b}$ also accounts for residual insertion or electronic losses not explicitly modeled.

\begin{figure*}[hb]
\centering
\includegraphics[width=\textwidth]{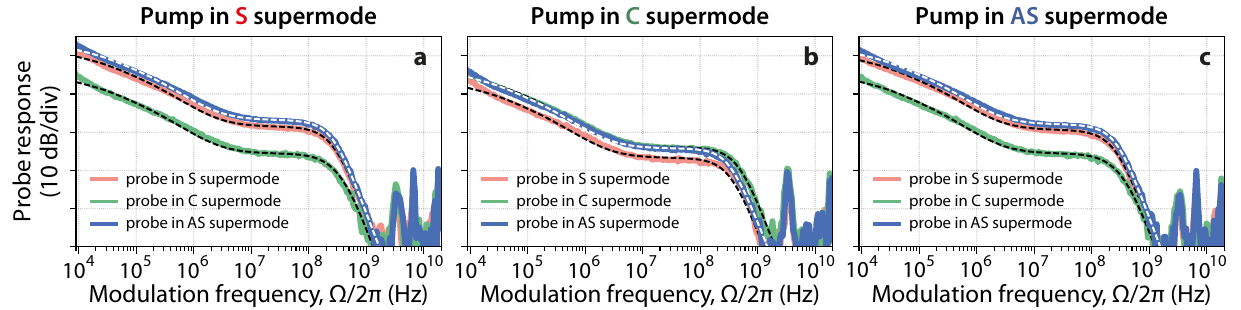}
\caption{\label{fig:overlapcharact}
Measured probe AC transmission as a function of the modulation frequency, with the pump fixed at the S \textbf{(a)}, C \textbf{(b)}, and AS \textbf{(c)}. Curves show probe response for S (red), C (green), and AS (blue) supermodes. Dashed lines are fits from \cref{eq:Tb_redefined}.
}
\end{figure*}

\cref{fig:overlapcharact} shows the measured probe response for all pump–probe combinations. When fitting \cref{eq:Tb_redefined}, $\kappa_{p(b)}$ and $\kappa_{p(b),\mathrm{ex}}$ are fixed using independent measurements, and only $N_{p,b}$, $X_{p,b}$, and $\Delta_b^{(0)}$ are fitted.
Interestingly, taking the product of $N_{p,b}$ and $X_{p,b}$, gives:
\begin{equation}
    N_{p,b} X_{p,b} = \frac{256 \omega_b c n_2 P_{in,b}}{n_0n_g}\Gamma_{p,b}.
\end{equation}
Therefore, by comparing this product across different pump–probe configurations, we directly measure the relative mode-overlap ratio:
\begin{equation}
    \frac{N_{p_1,b_1} X_{p_1,b_1}}{N_{p_2,b_2} X_{p_2,b_2}} \approx
    \frac{\Gamma_{p_1,b_1}}{\Gamma_{p_2,b_2}}.
\end{equation}

In the experiment, we fix the pump supermode and sweep through the three probe supermodes, repeating for each of the three pump modes. Since the pump wavelengths lie near the EDFA gain edge, each supermode experiences a different $P_{\mathrm{in},p}$. 
Therefore, we compare only across different probed supermodes for a fixed pump. This allows us to experimentally extract the relative overlap factors, as presented in Table 1 of the main text.

\bibliography{bib.bib}